\documentclass{jfm}
\usepackage{graphicx}
\usepackage{epstopdf, epsfig}
\usepackage{amssymb,amsmath,mathrsfs,amsbsy}
\usepackage{graphicx,multirow,soul,xcolor}

\shorttitle{Nonlinear evolution of the centrifugal instability using a semi-linear model}
\shortauthor{E. Yim, P. Billant and F. Gallaire}

\title{Nonlinear evolution of the centrifugal instability using a semi-linear model}

\author{Eunok Yim\aff{1}
  \corresp{\email{eunok.yim@epfl.ch}},
  P. Billant\aff{2}
 \and F. Gallaire\aff{1}}

\affiliation{\aff{1}LFMI, \'{E}cole Polytechnique F\'{e}d\'{e}rale de Lausanne, 1015 Lausanne,Switzerland
\aff{2}LadHyX, CNRS, \'{E}cole Polytechnique, F-91128 Palaiseau CEDEX, France}

\def\rt{\tilde{r}}
\def\rh{\hat{r}}
\def\be{\begin{equation}}
\def\ee{\end{equation}}

\def\i{\mathrm{i}}

\newcommand{\dab}[2]{\frac{\partial {#1}}{\partial {#2}}}

\begin{document}

\maketitle

\begin{abstract}
We study the nonlinear evolution of the centrifugal instability 
developing on a columnar anticyclone 
with a Gaussian angular velocity
using a semi-linear approach.
The model consists in two coupled equations: one for the linear evolution of the most unstable perturbation on the axially averaged mean flow
and another for the evolution of the mean flow under the effect of the axially averaged Reynolds stresses due to the perturbation.
Such model is similar to the self-consistent model of \cite{Vlado14} except that the time averaging is replaced by a spatial averaging.
The non-linear evolutions of the mean flow and the perturbations predicted by this semi-linear model are in very good agreement with DNS for 
the Rossby number $Ro=-4$ and both values of the Reynolds numbers investigated: $Re=800$ and $2000$ (based on the initial maximum angular velocity and radius of the vortex).
An improved model taking into account the second harmonic perturbations is also considered.
The results show that the angular momentum of the mean flow is homogenized towards a centrifugally stable profile
via the action of the Reynolds stresses of the fluctuations. 
The final velocity profile predicted by \cite{Kloosterziel07} in the inviscid limit is extended to finite high Reynolds numbers. It is in good agreement with 
the numerical simulations.
\end{abstract}

\begin{keywords}
Centrifugal instability, vortex, semi-linear model, stability analysis
\end{keywords}

\section{Introduction}
Centrifugal instability or inertial instability, is the most common instability developing on vortices in rotating medium. It is a local instability occurring when the balance between the centrifugal force and the pressure gradient is disrupted, i.e. when the square of the absolute angular momentum of the fluid decreases with radius $r$ in inviscid fluids \citep{Rayleigh17,Synge33,Kloosterziel91}. While this condition applies to axisymmetric disturbances, a generalized criterion for non-axisymmetric perturbations has been derived by \cite{Billant05a}.

Linear stability analysis of a columnar vortex with Gaussian angular velocity in inviscid fluids shows that the  growth rate is maximum at infinite wavenumber \citep{Smyth98}. 
However, as soon as viscous effects are taken into account, short wavelength  are damped and the fastest growing mode has a finite wavenumber \citep{Lazar13, Yim16}. 

\cite{Kloosterziel07} and \cite{Carnevale11} have analysed the nonlinear evolution of the centrifugal instability in a rotating medium at high Reynolds number. They have shown that the vortex saturates to a centrifugally stable state where the Rayleigh instability condition is no longer satisfied, i.e. the square of the axial average of the absolute angular momentum does not decrease with radius.
Hence, the instability redistributes the regions of positive and negative absolute angular momentum under the constraint of
absolute angular momentum conservation in the inviscid limit.

The saturation of an instability towards a periodic limit cycle for which the mean flow is stable has been recently described by means
of a self-consistent approach \citep{Vlado14}. In this approach, the flow is decomposed into time-averaged mean flow and unsteady perturbations. Then, 
the nonlinear saturation can be described by computing the mean flow distortion due to the Reynolds stresses of the perturbation and the linear
growth of the perturbation on the mean flow.
Here, we develop a similar approach for the centrifugal instability using a spatial average instead of a time average
since the instability is spatially periodic but not periodic in time.

\section{Governing equations}
We consider  a vortex
with angular velocity (figure \ref{fig:LISA}a)
\begin{equation}
\Omega = \Omega_0 \exp(-r^2/R^2), 
\label{eq:baseflow}
\end{equation}
 where $\Omega_0$ is the maximum angular velocity 
and $R$ the radius of the vortex. 
In the following, the length and time are non-dimensionalised with $R$ and $1/\Omega_0$, respectively. The governing equations for the velocity field $\mathbf{u}=[u,v,w]$ in cylindrical coordinates $(r,\theta, z)$ read 
\begin{align}
\frac{\partial {\mathbf{u}}}{\partial t} +  {\mathbf{u}} \cdot \nabla  {\mathbf{u}} + 2Ro^{-1}\mathbf{e}_z \times {\mathbf{u}} &=- \mathbf{\nabla}  {{p}} +Re^{-1} \nabla^2  {\mathbf{u}}, \label{eq:NS00} \\
\quad \nabla \cdot \mathbf{u} &= 0,
\label{eq:NS01}
\end{align}
where the Reynolds and Rossby numbers are defined as $Re=\Omega_0 R^2/\nu$ and $Ro=2\Omega_0/f$, respectively with $\nu$ the kinematic viscosity and $f$ the Coriolis parameter.

\subsection{Linear stability}
\begin{figure}
\centering
\centerline{\includegraphics[width=\textwidth]{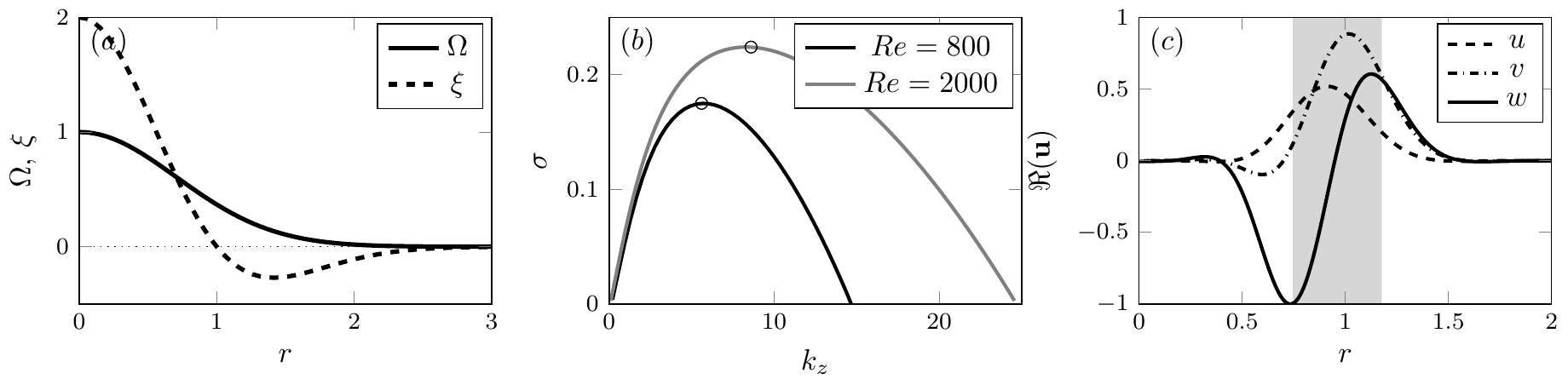}}
\caption{(a) 
Angular velocity ($\Omega$) and axial vorticity ($\xi$)
of the base flow. (b) Linear growth rate as a function of the vertical wavelength $k$ for $Re=800$ and $Re=2000$ for $Ro = -4$. (c) Real part of the most unstable eigenmode for $Re=800$. The shaded area indicates the region
where $\phi<0$.}
\label{fig:LISA}
\end{figure}

The linear stability of the base flow (\ref{eq:baseflow}) has been first studied by linearizing the equations (\ref{eq:NS00})--(\ref{eq:NS01})
and assuming axisymmetric infinitesimal perturbations with axial wavenumber $k$.
In the inviscid limit, the necessary and sufficient
condition for centrifugal instability reads \citep{Rayleigh17,Synge33,Kloosterziel91},
\begin{equation}
\phi = 2\left(\frac{v}{r} + \frac{1}{Ro} \right)\left({\xi}+\frac{2}{Ro} \right) <0,
\label{eq:phi}
\end{equation}
where  $\xi = 1/r \partial (r v)/\partial r$ is the axial vorticity.
For the base flow (\ref{eq:baseflow}), (\ref{eq:phi}) is satisfied when $Ro < -1$ and $Ro > \exp(2)$.
Figure \ref{fig:LISA}b shows the linear growth rate $\sigma$ as a function of $k$ for $Ro=-4$ for two different Reynolds numbers, $Re=800$ and $Re=2000$. 
The growth rate is maximum at $k_m = 5.6$ for $Re=800$ and $k_m=8.6$ for $Re=2000$.
 Figure \ref{fig:LISA}c shows the most unstable eigenmode for $Re=800$, $Ro=-4$.
It is mostly localized in the region where the
Rayleigh discriminant is negative (shaded area).

\section{Semi-linear formulation}
We decompose the flow as
\begin{equation}
\mathbf{u}(r,z,t) = \overline{\mathbf{u} }(r,t)  + \hat{\mathbf{u} }(r,z,t), 
\end{equation}
where $\overline{\mathbf{u} }=z_{\max}^{-1} \int_0^{z_{\max}} \mathbf{u}\mathrm{d} z$
is the axially averaged mean flow and $\hat{\mathbf{u} } $ the perturbation which is not assumed
to be small as in the linear stability analysis.
Averaging the equation (\ref{eq:NS00}) in $z$ leads to
\begin{equation}
\frac{\partial \overline{\mathbf{u}}}{\partial t} +  \overline{\mathbf{u}} \cdot \nabla  \overline{\mathbf{u}} + 2Ro^{-1}\mathbf{e}_z \times \overline{\mathbf{u}} + \mathbf{\nabla}  \overline{{p}} -Re^{-1} \nabla^2  \overline{\mathbf{u}} = - \overline{ \hat{\mathbf{u}} \cdot \nabla  \hat{\mathbf{u}}}. \label{eq:NSavg}
\end{equation}
Substracting (\ref{eq:NSavg}) from (\ref{eq:NS00}) yields
the equation for the perturbation $\hat{\mathbf{u} } $,
\begin{equation}
\frac{\partial  \hat{\mathbf{u}}}{\partial t}+  \overline{\mathbf{u}} \cdot \nabla  \hat{\mathbf{u}} + \hat{\mathbf{u}} \cdot \nabla  \overline{\mathbf{u}} + 2Ro^{-1}\mathbf{e}_z \times  \hat{\mathbf{u}} + \mathbf{\nabla} \hat{{p}} -Re^{-1} \nabla^2  \hat{\mathbf{u}} = -{ \hat{\mathbf{u}} \cdot \nabla  \hat{\mathbf{u}}}+\overline{ \hat{\mathbf{u}} \cdot \nabla  \hat{\mathbf{u}}}.\label{eq:NSpert}
\end{equation}

\subsection{Single harmonic}
We introduce now the normal mode form of the perturbation $\hat{\mathbf{u} }(r,z,t)  \simeq 
 {\mathbf{u} }_1(r,t) \mathrm{exp}(\mathrm{i} k_{m} z)+c.c.$ 
where $c.c$ indicates the complex conjugate and $k_{m}$ is the most amplified wavenumber obtained from the linear stability analysis. 
At $t=0$, the perturbation is set as ${\mathbf{u} }_1(r,0)=A_0\mathbf{u}_m$ where $\mathbf{u}_m$ is the dominant eigenmode and $A_0$ the initial amplitude of
the perturbation.
Neglecting the higher harmonics, the governing equations (\ref{eq:NSavg})-(\ref{eq:NSpert})  reduce to 
\begin{align}
%\mbox{Order} \ 0 \ :  \ 
\frac{\partial \overline{\mathbf{u}}}{\partial t}+\overline{\mathbf{u}} \cdot \nabla \overline{\mathbf{u}}+2Ro^{-1}\mathbf{e}_z \times  \overline{\mathbf{u}} + \nabla \overline{p} - Re^{-1} \nabla^2 \overline{\mathbf{u}}&=-  \zeta(\mathbf{u}_1), \label{eq:NSavgk} \\
%\mbox{Order } \exp(\mathrm{i} k_{m} z)  \ :  \qquad \qquad \qquad \qquad \qquad \qquad 
\frac{\partial \mathbf{u}_1 }{\partial t}  + \mathcal{L}(\overline{\mathbf{u}}){\mathbf{u}_1} &=0, \label{eq:NSpertk}
\end{align}
where $\mathcal{L}(\overline{\mathbf{u}})$ is a linear operator defined as $\mathcal{L}(\overline{\mathbf{u}})\hat{\mathbf{u}}    \equiv \overline{\mathbf{u}} \cdot \nabla_k  \hat{\mathbf{u}} + \hat{\mathbf{u}} \cdot \nabla  \overline{\mathbf{u}} + 2Ro^{-1} \mathbf{e}_z \times  \hat{\mathbf{u}} + \mathbf{\nabla}_k  \hat{{p}} -Re^{-1} \nabla^2_k  \hat{\mathbf{u}}$ and $\zeta(\mathbf{u}_1) = \mathbf{u}_1 \cdot \nabla_{-k} \mathbf{u}_1^* +  \mathbf{u}_1^* \cdot  \nabla_k  \mathbf{u}_1$ is the Reynolds stress. Here, $\nabla_k$ and $\nabla^2_k$ are respectively the gradient and Laplacian  in cylindrical coordinates with the vertical derivative replaced by $\i k$. 
It is worth mentioning that (\ref{eq:NSavgk}) -- (\ref{eq:NSpertk}) are now only a function of time and the radial coordinate $r$. In addition, 
the divergence free condition reduces to ${1}/{r} {\partial  r \overline{u}}/{\partial r}  = 0$ since ${\partial \overline{w}}/{\partial z} = 0$ due to the axial averaging.
This implies $\overline{u} = 0$.
It can also be shown that $\overline{w}$ remains identically zero for all time if $\overline{w}=0$ at $t=0$, since the Reynolds stress in the $z$ direction is zero. Thus, the mean flow has only a component along the azimutal direction $\overline{\mathbf{u}}=[0,\overline{v},0]^T$.
Hence, (\ref{eq:NSavgk}) simplifies to
\begin{equation}
\frac{\partial \overline{v}}{\partial t} = Re^{-1} \left[ \frac{\partial^2 \overline{v}}{\partial r^2} + \frac{1}{r} \frac{\partial \overline{v}}{\partial r} - \frac{\overline{v}}{r^2} \right] - \zeta(\mathbf{u}_1)_\theta ,
\label{eq:meansim}
\end{equation}
which is a simple diffusion equation with a source term independent from the Rossby number $Ro$.  

Therefore, our model consists in the semi-linear 1D equation (\ref{eq:NSpertk})
for the evolution of the perturbation over the mean flow coupled to the equation (\ref{eq:meansim}) for the evolution of the mean flow under the effects of the Reynolds stresses of the perturbation and viscous diffusion.
Such semi-linear model is similar to the self-consistent model proposed by \cite{Vlado14}. The main difference is that the Reynolds decomposition \citep{Reynolds72} to separate the flow into a mean flow and a fluctuation 
is here based on spatial axial average since the perturbation is harmonic along the axis while the self-consistent model relies upon a time average because the perturbation is harmonic in time for the flows they have considered.
Another difference is that the perturbation equations are here simply integrated in time while, in the self-consistent model, an eigenvalue
problem has to be solved after each variation of the mean flow.

\subsection{Two harmonics}
One can easily include higher harmonics of the fundamental mode following the same approach.
For instance, taking into account the second harmonic in the velocity perturbation:
$\hat{\mathbf{u}} =  {\mathbf{u} }_1(r) \mathrm{exp}(\mathrm{i} k_{m} z) + {\mathbf{u} }_2(r) \mathrm{exp}(\mathrm{i} 2 k_{m} z)+c.c.$, 
the perturbation equations become
\begin{subequations}
\begin{align}
%\mbox{Order } \exp(\mathrm{i} k_{m} z)  \ &: \ 
\frac{\partial \mathbf{u}_1 }{\partial t} +\mathcal{L}(\overline{\mathbf{u}}){\mathbf{u}_1}= - (\mathbf{{u}}_2 \cdot \nabla \mathbf{{u}}_1^*+\mathbf{{u}}_1^* \cdot \nabla \mathbf{u}_2), \\
%\mbox{Order } \exp(\mathrm{i} 2k_{m} z)  \ &: \ 
\frac{\partial \mathbf{{u}}_2}{\partial t}+ \mathcal{L}(\overline{\mathbf{u}}){\mathbf{{u}}_2}=-(\mathbf{{u}}_1 \cdot \nabla \mathbf{{u}_1}).
\end{align}
\label{eq:NSpertk2nd}
\end{subequations}
The mean flow (\ref{eq:meansim}) is then forced by the Reynolds stress of both harmonics
\begin{subequations}
\begin{align}
\frac{\partial \overline{v}}{\partial t} =Re^{-1} \left[ \frac{\partial^2 \overline{v}}{\partial r^2} + \frac{1}{r} \frac{\partial \overline{v}}{\partial r} - \frac{\overline{v}}{r^2} \right]  -  \zeta(\mathbf{u}_1)_\theta - \zeta(\mathbf{u}_2)_\theta.
\end{align}
\label{eq:meansin2nd}
\end{subequations}

Higher harmonics $3k_m, 4k_m, \cdots$
can be taken into account similarly. Since the perturbations equation are integrated in time, there is no particular complexity arising when several 
harmonics are considered. This is in contrast with the self-consistent model \citep{Vlado14} where the eigenvalue problems become increasingly complicated when more than one harmonic is considered \citep{Meliga17}.

\subsection{Numerical method}
All numerical simulations have been performed with the
FreeFEM++ software \citep{Hecht12}. The velocity and pressure are discretised with Taylor-Hood P2 and P1 elements, respectively.  
The time and the convection operators are discretised using characteristic-Garleken method for the direct numerical simulations (DNS) while the first order backward Euler formula is used 
for the semi-linear models. The total numbers of degree of freedom is $278817$ for the DNS and $4687$ for the semi-linear models, i.e. about 60 times less than for DNS.  
Both DNS and semi-linear models are initialized with 
the perturbation $\hat{\mathbf{u}}=A_0\mathbf{u}_m\mathrm{exp}(\mathrm{i} k_{m} z) + c.c. $ where $A_0=0.001$ is the initial amplitude and $\mathbf{u}_m$ is the most unstable linear eigenmode (figure \ref{fig:LISA}c) obtained by means of the restarted Arnoldi method.
The eigenmodes have been normalized so that the absolute maximum value of the vertical velocity is unity $\max (|w|)=1$.

The domain size is chosen to be $r=[0, \ r_{\max}]$ and $z=[0, \ z_{\max}]$ where $r_{\max} = 8$, $z_{\max}=2\pi/k_{m}$ and $k_{m}$ is the  most amplified axial wavenumber from the linear analysis. 
Periodic boundary conditions are applied on $z=0$ and $z=z_{\max}$. 
The boundary conditions at $r=0$ are $u = v =0$ since the flow is
axisymmetric \citep{Batchelor62}.
 At $r = r_{\max}$, all perturbations are enforced to vanish. 
Some DNS and simulations with the semi-linear models have been successfully checked against simulations with an independent pseudo-spectral code 
NS3D \citep{Deloncle08}.

\section{Results}
\subsection{DNS}
\begin{figure}
\centering
\centerline{\includegraphics[width=\textwidth]{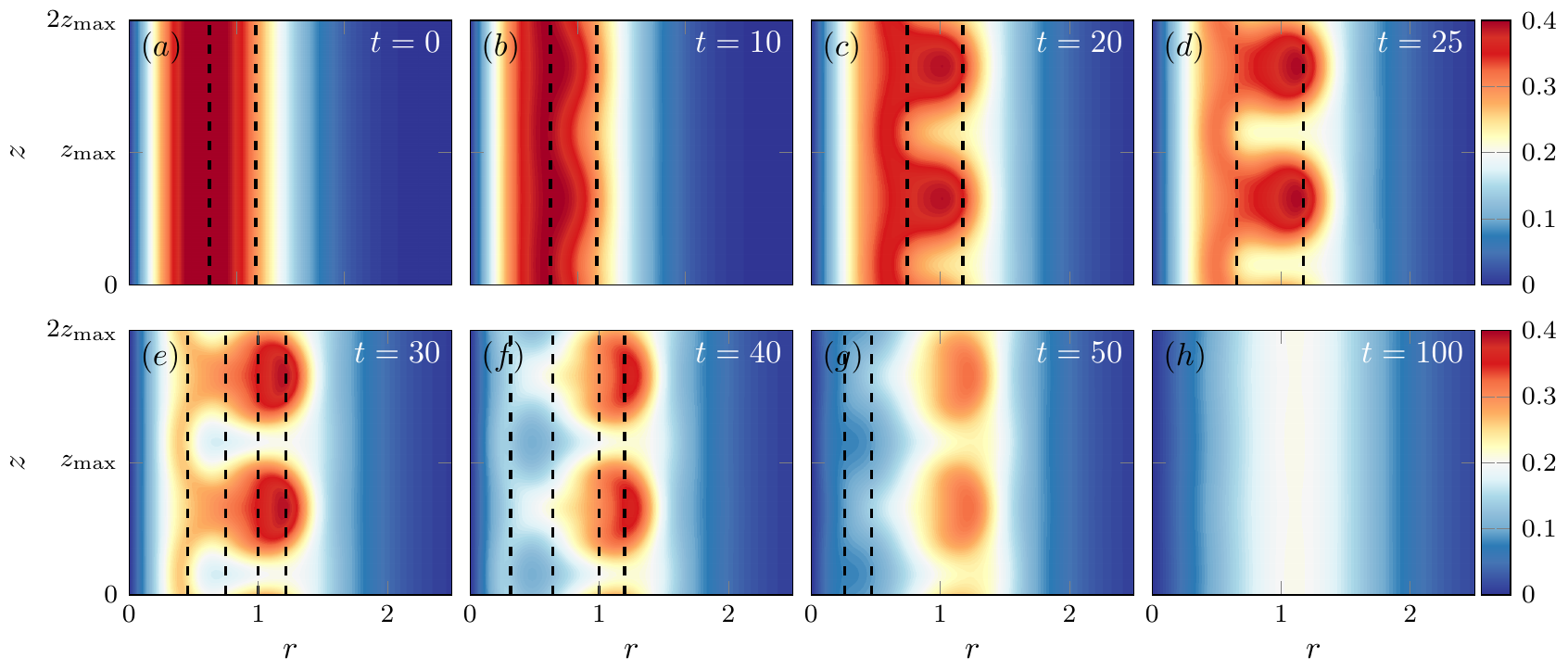}}
\caption{Evolution of the azimuthal velocity $v$ in DNS
for $k_{m}=5.6$, $z_{\max}=2 \pi/k_{m}$, $Ro = -4$, $Re=800$, $A_0 = 0.001$. The dashed lines delimit the regions where $\bar \phi < 0$ based on the mean azimuthal velocity $\bar{v}$.}
\label{fig:DNS800}
\end{figure}

Figure \ref{fig:DNS800} shows 
snapshots of the azimuthal velocity $v$ in a DNS for $Ro=-4$ and $Re=800$. Two wavelengths are displayed although the computation is performed over only one wavelength.
The vertical lines delimit the regions where the Rayleigh discriminant $\bar \phi$ is negative, where $\bar \phi$ is based on the axially averaged azimuthal velocity $\bar{v}(t,r)$.
At $t=10$, a slight deformation can be seen in the region where $\bar \phi < 0$. Subsequently, the perturbation grows and rearranges the distribution of azimuthal velocity ($20 < t< 30$).
For $t>30$, the `mushrooms' start to fade out.
Finally, vertical deformations are no longer visible by $t=100$ so that the vortex then evolves only by viscous diffusion.

The solid lines in figure \ref{fig:MeanVSC_800}a show the  evolution of the corresponding mean
azimuthal velocity $\bar{v}$.
The mean flow profile first decays by viscous diffusion
until $t=10$. A distortion of the mean flow due to the instability can be seen at
$t=20$.
At $t=30$ and $t=40$, it becomes strong and the profile exhibits 
two distinct peaks near $r=0.4$ and $r=1$. 
Then, the peak at $r\sim 0.4$ disappears and the mean azimuthal velocity profile becomes linear for $r < 1$ for $t>60$.
During this process, the maximum velocity has
decreased from $\max(\bar{v}(t=0))=0.47 $ to $\max(\bar{v}(t=80)) = 0.23 $ and the corresponding radius has moved from $r=0.7$ to $r=1.2$.
The corresponding Rayleigh discriminant 
is shown in figure \ref{fig:MeanVSC_800}b (solid lines). At $t=0$, 
$\bar \phi$ is minimum at $r=0.95$ and is negative  for $0.75 < r < 1.18$. For $t=20$, the region where $\bar \phi < 0$ has enlarged but the minimum of $\bar \phi$ has decreased in absolute value.
At $t=30$, there exist two 
regions where $\bar \phi<0$: near $r=0.3$ and $r=0.8$
while $\bar \phi > 0$ in between.
The minimum value decreases and
then increases again for larger time. At $t=60$, 
$\min(\bar \phi)$ is no longer negative.

 The evolution of the mean azimuthal velocity $\bar{v}$ is qualitatively 
similar for the higher Reynolds number $Re=2000$ (solid lines in figure \ref{fig:MeanVA_2000_k}a).  

\subsection{SL-1$k_m$ semi-linear model}
\begin{figure}
\centering
\centerline{\includegraphics[width=\textwidth]{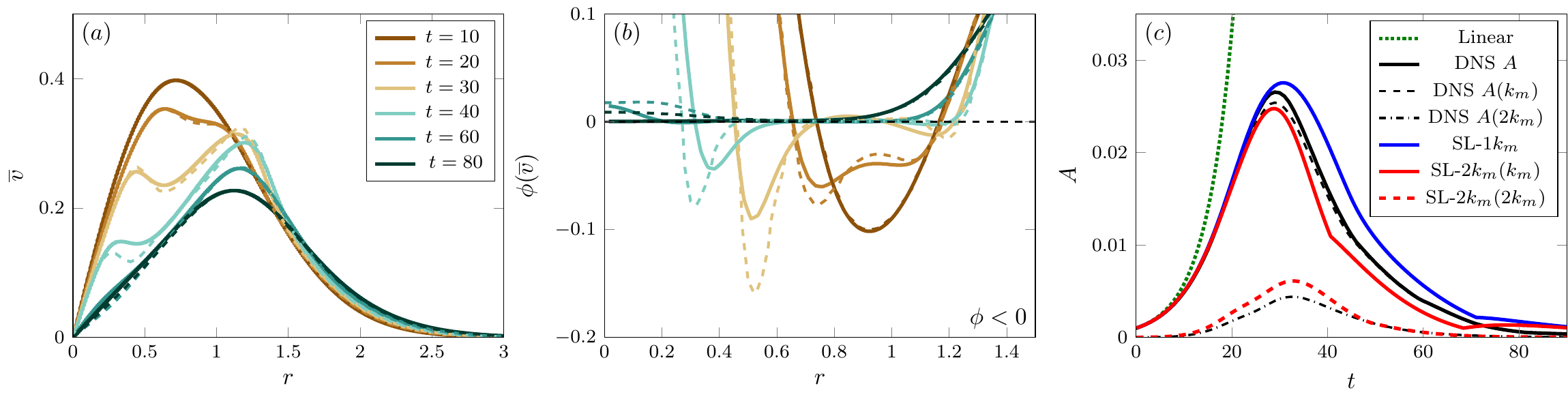}}
\caption{(a) Mean azimuthal velocity $\bar{v}$ from DNS (solid lines) and SL-1$k_m$ model (dashed lines with symbols), (b) corresponding Rayleigh discriminant $\bar \phi$ and (c)  perturbation amplitudes $A$ as a function of time for $Ro=-4$ and $Re=800$.}
\label{fig:MeanVSC_800}
\end{figure}

The lines with symbols in figure \ref{fig:MeanVSC_800}a represent the mean azimuthal flow predicted by the semi-linear model with one harmonic, abbreviated as SL-1$k_m$, for $Re=800$. The agreement with the DNS 
is almost perfect for $t=10, 20$ while some discrepancies can be seen at $t=30,40$.
It becomes excellent again for $t > 60$. Similar agreement and discrepancies are also observed for the Rayleigh discriminant $\bar{\phi}$ (figure \ref{fig:MeanVSC_800}b).   
Figure \ref{fig:MeanVSC_800}c shows the amplitude of the velocity perturbation in the DNS and the semi-linear model. The amplitude first increases exponentially from its initial value $A_0=0.001$.  
The linear prediction  (dotted line) 
agrees with the DNS only for small time ($t<10$). Then, the growth rate 
becomes smaller than the linear growth rate.
The perturbation grows until $t=30$ and then
decreases. 
The amplitude of the first and second harmonics $A(k_m)$, $A(2k_m)$ have been decomposed using FFT (broken lines).
The amplitude of the second harmonic is less than $20\%$ of the amplitude of the first harmonic. 
 The amplitude of the perturbation in the semi-linear model SL-1$k_m$ (blue thick line) is in very good agreement with the one in the DNS until $t=30$. Then, it slightly overestimates the amplitude extracted from the DNS. 

%The evolution of the total kinetic energy per unit vertical length
%\begin{equation}
%E_k(t) = \frac{1}{2 z_{\max}} \int_{0}^{z_{\max}} \int_{0}^{r_{\max}} (u^2+v^2+w^2) r \mathrm{d}r \mathrm{d}z 
%\end{equation}
%in the DNS and SL-1$k_m$ model are also in very good agreement (figure \ref{fig:KE}a).

 For the higher Reynolds number $Re=2000$,
the evolution of the mean azimuthal flow and amplitudes of the perturbation in the DNS and SL-$1k_m$ model are also globally in good agreement (figure \ref{fig:MeanVA_2000_k}a,c).
Nevertheless, some deviations can be seen at $t=30, 40$, in particular the first peak at small radius is not captured at $t=40$.
In addition, the maximum amplitude of the second harmonic reaches  $1/3$ of the maximum amplitude of the fundamental harmonic (figure \ref{fig:MeanVA_2000_k}c).
%There is a slight discrepancy between the evolutions of the kinetic energy in the DNS and SL-$1k_m$ model (figure \ref{fig:KE}b). 

\subsection{SL-2$k_m$ semi-linear model}
  \begin{figure}
\centering
\centerline{\includegraphics[width=\textwidth]{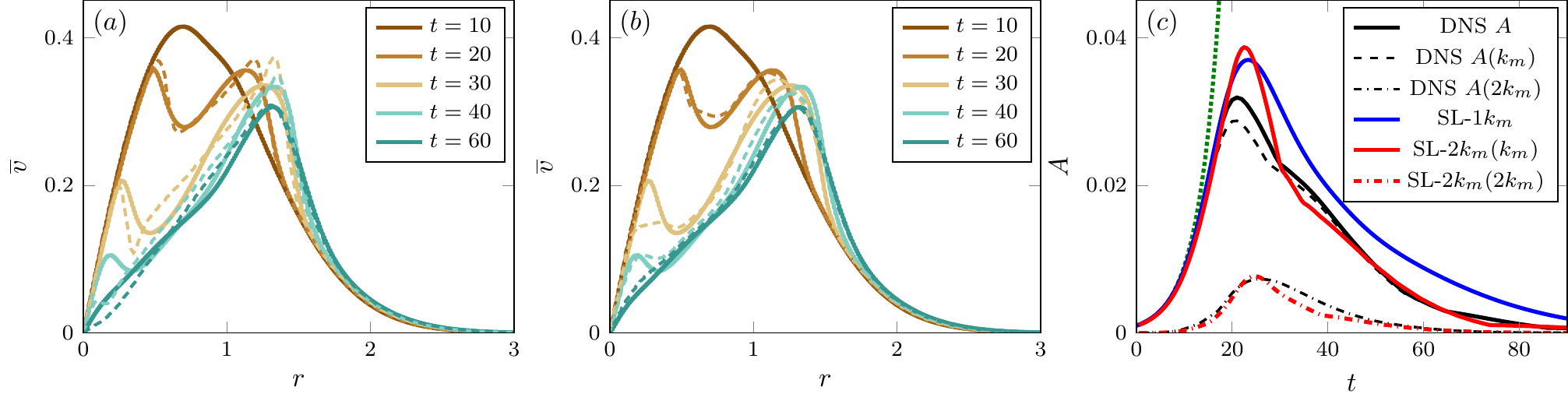}}
\caption{ (a) Mean azimuthal velocity
from DNS (solid lines)
and semi-linear model SL-$1k_m$
(symbols)
and (b) for SL-2$k_m$. (c) The perturbation amplitudes $A$ as a function of time for $Ro=-4$ and $Re=2000$ for both SL-$1k_m$ and SL-2$k_m$. }
\label{fig:MeanVA_2000_k}
\end{figure}
%
%\begin{figure}
%\centering
%\centerline{\includegraphics[width=0.8\textwidth]{MeanVA_2000.pdf}}
%\caption{Same as figure \ref{fig:MeanVA_2000_k} but for the SL-2$k_m$ model.}
%\label{fig:MeanVPhi_Re2000}
%\end{figure}
% 

\begin{figure}
\centering
\centerline{\includegraphics[width=\textwidth]{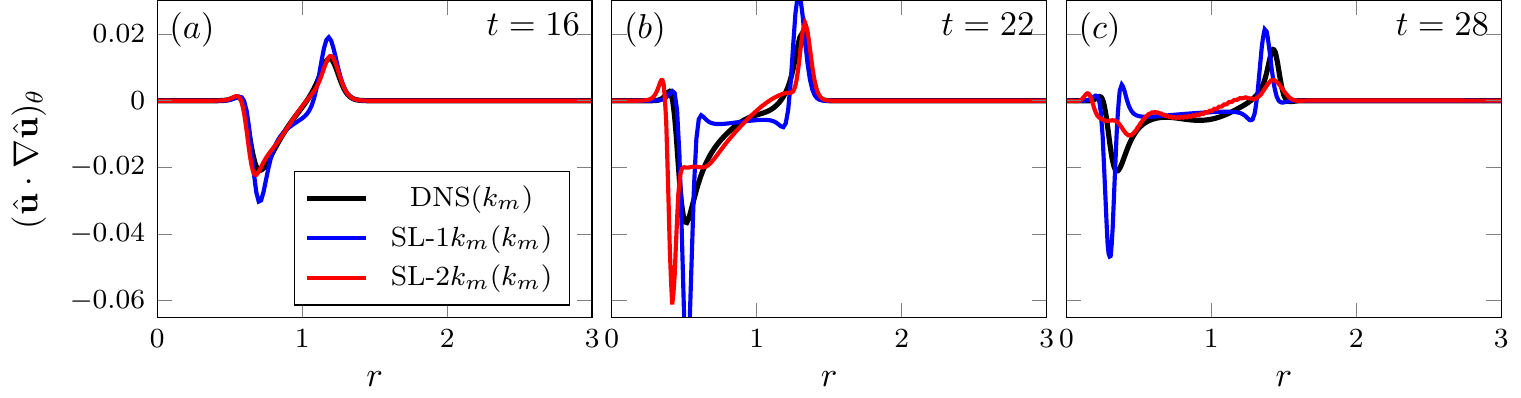}}
%ReystressTh2000_sameAxist102030.pdf}}
\caption{
Mean Reynolds stresses in the azimuthal direction from DNS and semi-linear models at
(a) $t=16$, (b) $t=22$ and (c) $t=28$. }
\label{fig:ReystressTh2000}
\end{figure}
As seen in figures \ref{fig:MeanVSC_800}c and \ref{fig:MeanVA_2000_k}c, the second harmonic is triggered and reaches a non-negligible amplitude for
both $Re=800$ and $Re=2000$. Its effect can be taken into account by means of the semi-linear model (\ref{eq:NSpertk2nd}-\ref{eq:meansin2nd}), called SL-$2k_m$. Since the perturbation is initialized by only the leading eigenmode, the initial amplitude of the
second harmonic is set to zero. Therefore, even if it is also unstable for $Re=2000$ (figure \ref{fig:LISA}b), its initial evolution is only due to the forcing by the fundamental harmonic.

 For $Re=800$ (figure \ref{fig:MeanVSC_800}c), the evolution of the amplitude of the first harmonic predicted by the SL-$2k_m$ model is closer to the DNS than for the SL-$1k_m$ model.
The amplitude of the second harmonic is also well predicted by the SL-$2k_m$ model.

 For $Re=2000$ (figure \ref{fig:MeanVA_2000_k}c), the 
predicted profiles for the mean azimuthal velocity are smoother than for the SL-$1k_m$ model (figure \ref{fig:MeanVA_2000_k}a). 
However, the first peak near $r\sim 0.3$ at $t=30$ is not captured. 
To understand this discrepancy, we have plotted the Reynolds stresses in the $\theta$-direction in the DNS and the semi-linear models (figure \ref{fig:ReystressTh2000}). 

At $t=16$, SL-$2k_m$ are in excellent agreement with DNS while at $t=22$ and $t=28$, the SL-$2k_m$ model agrees generally better with the DNS except the first minimum at small radius at $t=22$. This explains why the SL-$2k_m$ model does not capture the first peak of the mean azimuthal velocity (figure \ref{fig:MeanVA_2000_k}b). 

%At $t=10$, both models SL-$1k_m$ and SL-$2k_m$ are in good agreement with DNS while at $t=20$ and $t=30$, the SL-$2k_m$ model agrees generally better with the DNS except the first minimum at small radius at $t=30$. This explains why the SL-$2k_m$ model does not capture the first peak of the mean azimuthal velocity (figure \ref{fig:MeanVA_2000_k}b). 

Of course, including higher harmonics should further improve the predictions.
 However, this would complexify the models while the primary goal
of the present approach is the simplicity rather than the accuracy. 

\begin{figure}
\centering
%\centerline{\includegraphics[width=\textwidth]{Kinetic_Energy_finalprofile2.pdf}}
%\caption{(a,b) Total kinetic energy for (a) $Re=800$ and (b) $Re=2000$.  
% Viscous-KCO indicates the kinetic energy of the theoretical final velocity profile (\ref{eq:asympOm}) which takes into account viscous effects. (c) Mean azimuthal velocity in the DNS and semi-linear models at $t=60$ 
%for $Ro=-4$ for  $Re=800$ and $Re=2000$. 
%KCO indicates the velocity distribution predicted by  \cite{Kloosterziel07}  in the inviscid limit. 
%Viscous-KCO corresponds to the velocity profile (\ref{eq:asympOm}) which takes into account viscous effects.}
\centerline{\includegraphics[width=0.75\textwidth]{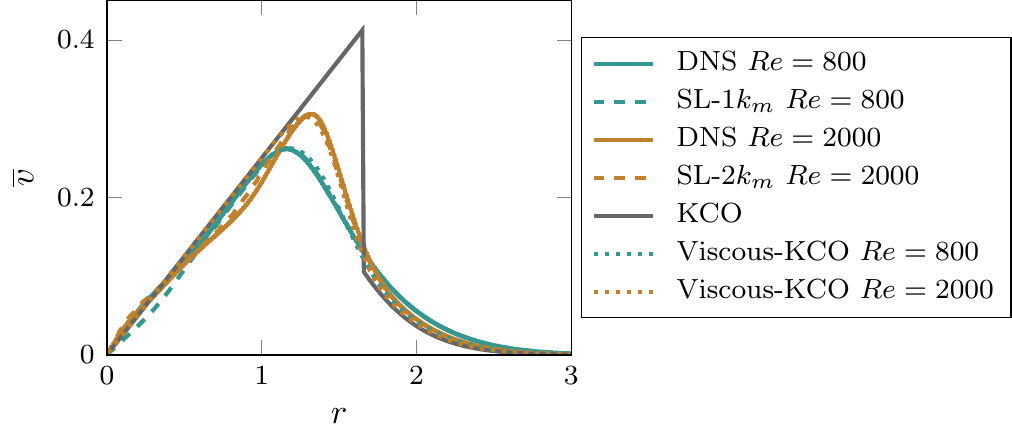}}
\caption{ Mean azimuthal velocity in the DNS and semi-linear models at $t=60$ 
for $Ro=-4$ for  $Re=800$ and $Re=2000$. 
KCO indicates the velocity distribution predicted by  \cite{Kloosterziel07}  in the inviscid limit. 
Viscous-KCO corresponds to the velocity profile (\ref{eq:asympOm}) which takes into account viscous effects.}
\label{fig:KE}
\end{figure}
 
\section{Final profiles of azimuthal velocity}
The profiles of mean azimuthal velocity observed at late time $t=60$, once the instability has ceased, have been compared to the
theory of \cite{Kloosterziel07} and \cite{Carnevale11}. As mentioned in the introduction, this 
theory states that the centrifugal instability homogenizes negative and positive absolute angular momentum $L=r(v + r/Ro)$
so as to suppress negative gradients of $L^2$ under the constraint of absolute angular momentum conservation. 
For anticyclones, the final profile of $L$ in the inviscid limit is such that
$L$ is zero until a radius $r_c$ given by $\int_0^{r_c} rL_i \rm{d}r=0, $
%\begin{equation}
%\int_0^{r_c} rL_i \rm{d}r=0, 
%\label{eq:tiaam}
%\end{equation}
where $L_i$ is the initial absolute angular momentum.
Beyond this radius, the
velocity profile remains identical to the initial one. Therefore, the theoretical angular velocity is
\begin{equation}
\Omega(r<r_c) = -\frac{1}{Ro}, \qquad \Omega(r \geq r_c) = \exp(-r^2).
\label{eq:invpro}
\end{equation}
This profile (labelled KCO) is compared to those observed in the DNS and SL models for $Re=800$ and $Re=2000$ in figure \ref{fig:KE}.
It is close to the observed profiles except in the vicinity of the radius $r_c$ where the latter are smooth while the theoretical profile 
is discontinuous due to the inviscid approximation.
In order to take into account viscous effects, we have further considered the viscous diffusion of (\ref{eq:invpro}). For large Reynolds number, the diffusion equation 
\begin{equation}
\frac{\partial \Omega}{\partial t} = Re^{-1} \left[ \frac{\partial^2 \Omega}{\partial r^2} + \frac{3}{r} \frac{\partial \Omega}{\partial r} \right].
\label{eq:apom}
\end{equation}
shows that the angular velocity should decay slowly everywhere except in the vicinity of $r_c$ where
radial derivatives are expected to be large because of the discontinuity. To describe the local viscous evolution near $r_c$, we therefore define a rescaled radial coordinate $\rt =\sqrt{Re} (r-r_c)$ and we assume that $\Omega$ depends both on $\rt$ and the unscaled radius $\rh =r$.
We also introduce a slow time $\tau= Re^{-1} t$.
Hence, (\ref{eq:apom}) becomes
\begin{equation}
\frac{\partial \Omega}{\partial t} + \frac{1}{Re}\frac{\partial \Omega}{\partial \tau}= \frac{\partial^2 \Omega}{\partial \rt^2} + \frac{1}{\sqrt{Re}} \left(2 \frac{\partial^2 \Omega}{\partial \rt \partial \rh} + \frac{3}{\rh} \frac{\partial \Omega}{\partial \rt}\right) + \frac{1}{Re} \left(\frac{\partial^2 \Omega}{\partial \rh^2} + \frac{3}{\rh} \frac{\partial \Omega}{\partial \rh} \right). 
\label{eq:apom2}
\end{equation}
Then, the solution is sought as an expansion in Reynolds number
\begin{equation}
\Omega = \Omega_0 + {Re}^{-1/2}\Omega_1 + Re^{-1}\Omega_2 + \cdots.
\end{equation}
The zero-th and first order problems are 
\begin{equation}
\frac{\partial \Omega_0}{\partial t} = \frac{\partial^2 \Omega_0}{\partial \rt^2}, \quad\mbox{and} \quad \dab{\Omega_1}{t}-\dab{^2 \Omega_1}{\rt^2} = 2 \frac{\partial^2 \Omega_0}{\partial \rt \partial \rh} + \frac{3}{\rh} \frac{\partial \Omega_0}{\partial \rt}.
\end{equation} 
The solutions are chosen as
\begin{equation}
\Omega_0 = A(\rh,\tau){\rm erf} \left( \frac{\rt}{2 \sqrt{t}} \right)+H(\rh,\tau), \quad\mbox{and} \quad \Omega_1=\left(2 \dab{A}{\rh}+3 \frac{A}{\rh} \right) \sqrt{\frac{t}{\pi}}\exp\left(- \frac{\rt^2}{4t} \right).
\end{equation}
where $A$ and $H$ are arbitrary functions of $\rh$ and $\tau$. These functions are found by considering the problem at order $Re^{-1}$: 
\begin{equation}
\dab{\Omega_2}{t}-\dab{^2 \Omega_2}{\rt^2} = -\dab{\Omega_0}{\tau} + \dab{^2 \Omega_0}{\rh^2}+\frac{3}{\rh} \dab{\Omega_0}{\rh} + 2 \dab{^2 \Omega_1}{\rh \partial \rt} + \frac{3}{\rh}\dab{\Omega_1}{\rt}.
\label{eq:appom2}
\end{equation}
It can be shown that the solution $\Omega_2$ presents secular growth unless we set
\begin{equation}
\dab{A}{\tau} = \dab{^2 A}{\rh^2} + \frac{3}{\rh}\dab{A}{\rh},\qquad\mbox{and} \qquad
\dab{H}{\tau} = \dab{^2 H}{\rh^2} + \frac{3}{\rh} \dab{H}{\rh}.
\end{equation}
The solutions are taken as 
\begin{equation}
A=\frac{D \exp\left(-\frac{\rh^2}{B+4\tau} \right)}{(B+4\tau)^2} + C,\qquad\mbox{and} \qquad H=\frac{E \exp\left(-\frac{\rh^2}{G+4\tau} \right)}{(G+4\tau)^2} +F,
\label{eq:AH}
\end{equation}
where $B, C$, $D$, $E$, $F$ and $G$ are constants. 
We then impose that $\Omega_0$ at $t=\tau=0$ matches the profile (\ref{eq:invpro}). 
This implies $B=G=1$, $E=D=1/2$, $F=-C=-1/(2Ro)$. Then, the solution of (\ref{eq:appom2}) can be found
\begin{align}
\Omega_{2}&= -\frac{1}{\sqrt{\pi}}\left(\dab{^2 A}{r^2}+\frac{3}{r} \dab{A}{r} +\frac{3}{4r^2} A \right) \rt \sqrt{t} \exp\left(- \frac{\rt^2}{4t} \right).
\end{align}
Finally, the complete solution for $\Omega$ up to order ${Re^{-1}}$, written back in terms of the original variables $r$ and $t$, reads
\begin{align}
\Omega &=  A {\rm erf}\left( \frac{\sqrt{Re} (r-r_c)}{2 \sqrt{t}} \right) + A -\frac{1}{Ro} + \left[2 \dab{A}{r}+ \frac{3}{r}A  \right.\nonumber\\
&\left. - \left(\dab{^2 A}{r^2}+\frac{3}{r} \dab{A}{r} +\frac{3}{4r^2} A \right)(r-r_c)\right]
\sqrt{\frac{t}{\pi Re}} \exp\left(-\frac{Re(r-r_c)^2}{4t}\right), 
%\Omega &=  A {\rm erf}\left( \frac{\sqrt{Re} (r-r_c)}{2 \sqrt{t}} \right) + A -\frac{1}{Ro} \nonumber\\
%+& \left[2 \dab{A}{r}+ \frac{3}{r}A  - \left(\dab{^2 A}{r^2}+\frac{3}{r} \dab{A}{r} +\frac{3}{4r^2} A \right)(r-r_c)\right]
%\sqrt{\frac{t}{\pi Re}} \exp\left(-\frac{Re(r-r_c)^2}{4t}\right), 
\label{eq:asympOm}
\end{align}
where $A$ is given by (\ref{eq:AH}) with the substitution $\rh =r$ and $\tau= Re^{-1} t$.
The azimuthal velocity profiles corresponding to (\ref{eq:asympOm}) are plotted with dotted lines (viscous-KCO) in figure \ref{fig:KE}.
The time in (\ref{eq:asympOm}) has not been set to $t=60$ but to $t=20$.
Indeed, since the profile (\ref{eq:invpro}) is the outcome of the centrifugal instability, we have assumed that it is virtually formed only at $t=40$ once the instability has almost ceased.
These viscous profiles are in much better agreement with the DNS than the inviscid profile of \cite{Kloosterziel07} and \cite{Carnevale11}.
Besides, we emphasize that the profiles in the DNS and SL models are in excellent agreement.

\section{Conclusion}
We have studied the nonlinear growth of the centrifugal instability in an anticyclone
with Gaussian angular velocity in rotating fluids for $Ro=-4$.
We  have used an approach similar to the one behind the self-consistent model \citep{Vlado14}. 
Using Reynolds decompositions \citep{Reynolds72} based upon a time average, \cite{Vlado14} have separated the flow into mean flow and time harmonic fluctuations. 
These two components are coupled via Reynolds stresses in the mean flow equation and via the evolution of the mean flow in the fluctuation equation. 
In the present study on the centrifugal instability, we have used a spatial average instead of a time average and separated the flow into axially averaged mean flow and spatial harmonic fluctuation. 
Like for the self-consistent model, the fluctuation grows over an evolving mean flow while the mean flow is forced by the Reynolds stresses due to the fluctuations.
Such semi-linear model with one harmonic is in very good agreement with DNS for $Re=800$ and $Re=2000$ concerning
both the time evolution of the fluctuation amplitude and of the mean flow profiles. 
Including a second harmonic $2k_m$ into the model improves 
slightly the predictions.

We have also compared the `final' azimuthal velocity profile 
observed in the DNS and semi-linear models when the instability has disappeared to the inviscid
profile proposed by \cite{Kloosterziel07} based on
homogenization of angular momentum 
towards a centrifugally stable flow. They agree except in the neighbourhood of the radius where the inviscid profile is discontinuous. 
To improve the prediction, we have computed asymptotically for large Reynolds number the viscous diffusion of the theoretical
profile of \cite{Kloosterziel07}. The discontinuity is then smoothed and the predicted profiles are in much better agreement with the profile observed in the DNS and semi-linear models.

The main interest of the present semi-linear models is their simplicity which may enable a deeper understanding of the underlying physics. In addition, we emphasize that they are very cheap in terms of 
computational cost. Indeed, the computing time for a DNS with four processors 
takes 30 hours (elapsed real time) for 100 time units. In contrast, a run of the semi-linear model with one or two harmonics only takes 6min or 10min, respectively with a single processor.
This dramatic decrease on the computing time comes from the reduction of the problem to only few one-dimensional equations. 

In the future, it would be interesting to investigate the connection between semi-linear models and amplitude equations derived from weakly nonlinear analyses.

%In the future, it would be interesting to consider similar models for the non-axisymmetric centrifugal instability since it can be dominant in presence of background stratification \citep{Lahaye15,Yim16, Yim19}.  
%A similar approach could be also conducted for the two-dimensional shear instability by means of an azimuthal average.

Declaration of Interests. The authors report no conflict of interest.

\bibliographystyle{jfm}
\bibliography{biblio_all}

\end{document}